\documentclass[journal,twoside,web]{ieeecolor}
\usepackage{jsen}
\usepackage{cite}
\usepackage{amsmath,amssymb,amsfonts}
\usepackage{graphicx}
\usepackage{textcomp}
\usepackage{wrapfig}
\usepackage{amsmath,amssymb,amsfonts}
\usepackage{algorithmicx}
\usepackage{algorithm}
\usepackage{graphicx}
\usepackage{textcomp}
\usepackage{textcomp}
\usepackage{graphicx}
\usepackage{subfigure} 
\usepackage{url}
\usepackage{verbatim}
\usepackage{graphicx}
\usepackage{cite}
\usepackage{CJK}
\usepackage{soul}
\usepackage{multirow}
\usepackage{threeparttable}
\usepackage{varwidth}
\usepackage{booktabs}
\usepackage{amsmath,bm}\usepackage{threeparttable}
\usepackage{varwidth}
\usepackage{booktabs}
\usepackage{amsmath,bm}
\usepackage{color}
\usepackage{algpseudocode}
\usepackage{makecell}
\makeatletter
\newenvironment{breakablealgorithm}
{
		\begin{center}
			\refstepcounter{algorithm}
			\hrule height.8pt depth0pt \kern2pt
			\renewcommand{\caption}[2][\relax]{
				{\raggedright\textbf{\ALG@name~\thealgorithm} ##2\par}%
				\ifx\relax##1\relax 
				\addcontentsline{loa}{algorithm}{\protect\numberline{\thealgorithm}##2}%
				\else 
				\addcontentsline{loa}{algorithm}{\protect\numberline{\thealgorithm}##1}%
				\fi
				\kern2pt\hrule\kern2pt
			}
		}{
		\kern2pt\hrule\relax
	\end{center}
}
\makeatother
\def\BibTeX{{\rm B\kern-.05em{\sc i\kern-.025em b}\kern-.08em
    T\kern-.1667em\lower.7ex\hbox{E}\kern-.125emX}}
\definecolor{abstractbg}{rgb}{0.89804,0.94510,0.83137}
\begin{document}
\title{Estimating Continuous Muscle Fatigue For Multi-Muscle Coordinated Exercise: A Pilot Study on Walking}

\author{Chunzhi Yi, \IEEEmembership{Member, IEEE}, Xiaolei Sun, Chunyu Zhang, Wei Jin, Jianfei Zhu, Haiqi Zhu, Baichun Wei
\thanks{C. Yi, H. Zhu and B. Wei are with the School of Medicine and Health, Harbin Institute 
of Technology, Harbin 150000, Heilongjiang, China.}
\thanks{C. Yi and W. Jin are with Zhengzhou Research Institute of Harbin Institute of Technology, Zhengzhou 450000, Henan, China.}
\thanks{X. Sun is with the Graduate School, Harbin Sport University, Harbin 150000, China and the    Department of Rehabilitation Medicine, the Fifth Hospital of Harbin City, Harbin 150000, China.}
\thanks{C. Zhang is with the School of Sport Medicine and Rehabilitation, Beijing Sport University, Beijing 100084, China.}
\thanks{J. Zhu is with the Faculty of Computing, Harbin Institute of Technology, Harbin 150000, Heilongjiang, China.}
\thanks{C. Yi and X. Sun share the same contribution.}
\thanks{Corresponding author: Haiqi Zhu, Baichun Wei (email:  haiqizhu@hit.edu.cn, bcwei@hit.edu.cn).}
}

\IEEEtitleabstractindextext{%
\fcolorbox{abstractbg}{abstractbg}{%
\begin{minipage}{\textwidth}%
\begin{wrapfigure}[16]{r}{3.6in}%
	\includegraphics[width=3.5in]{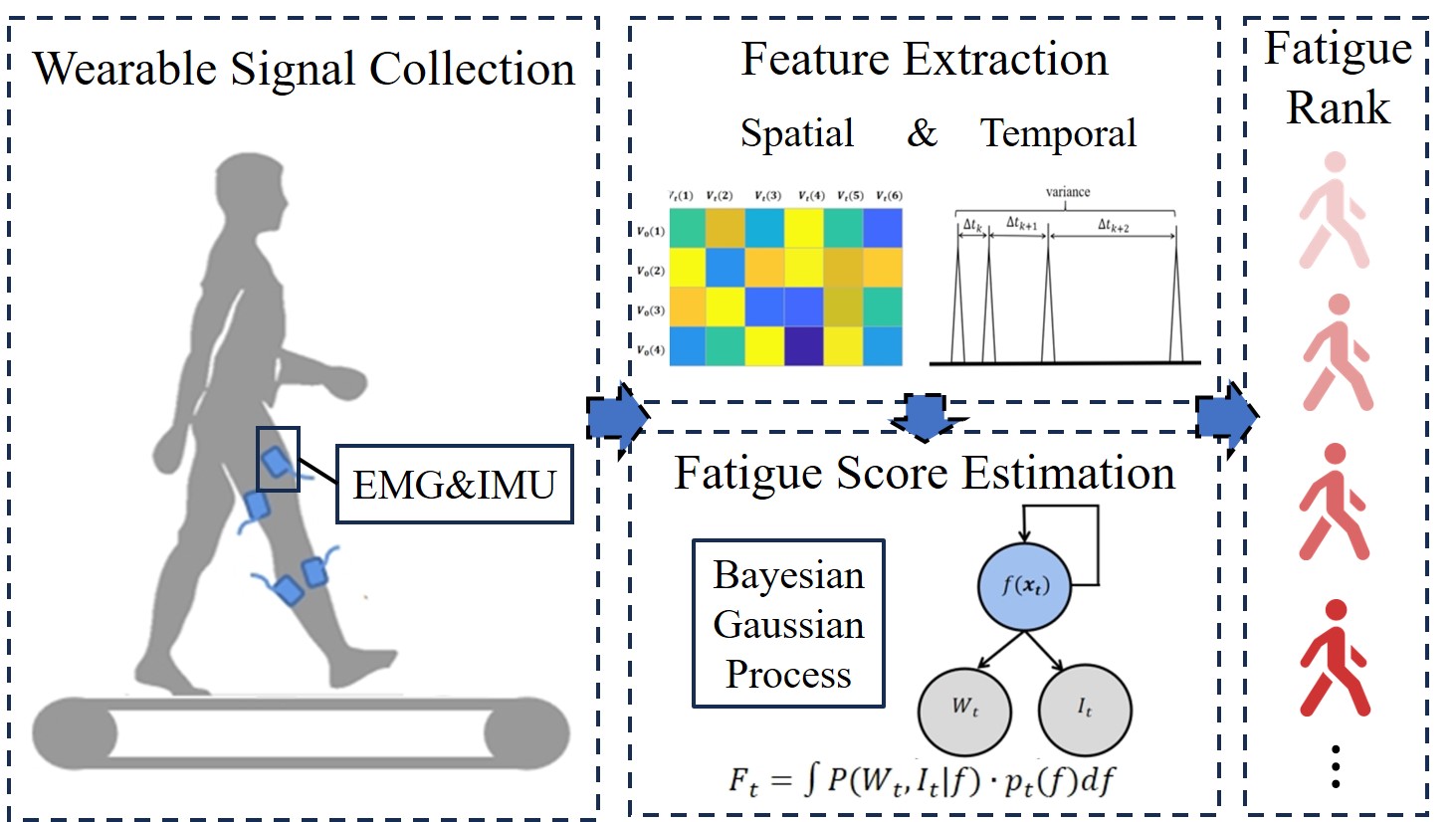}%
\end{wrapfigure}%

\begin{abstract}
Continuously assessing the progression of muscle fatigue for daily exercises provides vital indicators for precise rehabilitation and personalized training. However, current methods are limited in predicting the group-averaged progression of muscle fatigue, cannot esimate the individual muscle fatigue continuously. In this paper, we propose to depict fatigue by the features of  muscle compensation and spinal module activation changes  and estimate continuous fatigue by a physiological rationale model. First, we extract muscle synergy fractionation and the variance of spinal module spikings as features inspired by fatigue-induced neuromuscular adaptations during fatigue. Second, we formulate the fatigue estimator by developing a Bayesian Gaussian process to capture the time-evolving progression. Third, we train the algorithm by developing a weakly monotonically increasing function as the loss function inspired by the time-evolving characteristics of fatigue. Finally, we adapt the metrics that follow the physiological principles of fatigue to quantitatively evaluate the performance. We evaluate our algorithm by extensive experiments on weak monotonicity, cross-day, cross subject and  cross-view similarities.  This study would aim the objective assessment of muscle fatigue. 
\end{abstract}

\end{minipage}}}

\maketitle

\section{Introduction}
\label{sec:introduction}
\IEEEPARstart{M}{uscle} fatigue relates to the ability reduction of generating force or power after repetitive and intensive exercise \cite{enoka2008muscle}. Assessing muscle fatigue is of significant importance for sports, rehabilitation, human-robot interaction, even medical diagnosis  \cite{liu2019emg,gruet2018fatigue,peternel2019selective,yancey2012chronic}. Other than discretely classifying muscles are fatigued or not \cite{marco2017surface,zhang2018non,mugnosso2018muscle}, depicting the continuous progression of muscle fatigue during daily exercise enables assessing to what extent muscles are fatigued, and would  provide a physiological indicator for remote and interactive rehabilitative training and aids physicians' more fine-grained determination of dose and treatment.  Current  methods for continuously assessing muscle fatigue focus on isometric and isokinetic contractions \cite{xu2018fatigue, rocha2017weighted}, the endurance time of specific physical tasks \cite{drake2024modelling, liu2018experimental}, or the physiological modelling of fatigue mechanism \cite{jebelli2020physiology, FREYLAW2021103104, frey2019optimisation}. How to use physiological measures to continuously assess the muscle fatigue of daily exercise (e.g.  running, walking or bycycling) in real time is still not fully explored.

To do so, it requires to extract fatigue-related features from physiological measures or signals and to estimate a continuous score of muscle fatigue. On feature extraction, extensive studies focus on noninvasive signal-extracted features that discretely reflect fatigue-induced changes of local muscle properties, which outperform the biomarker-based features (e.g. evoked potential-based features using stimulation  \cite{gruet2013stimulation} , the biochemical changes  \cite{siegel2008cardiac} or the the decline of motor performance \cite{mugnosso2018muscle,sanchez2011velocity}) in real-time sensing.  Fatigued muscles present changes of multiple physiological properties including lactic acid, blood oxygen metabolism, motor unit recruitment, and muscle fiber conduction velocity. Inspired by the phenomena of low frequency shift and increased amplitude, classical features like root mean square, median frequency and mean power frequency etc. are used to present temporal and spectral changes of electromyography (EMG) signals \cite{zhang2018non,wang2021novel}. High-density EMG decomposition further enables the detection of subtle muscle fibre properties, thus provides more precise features of individual muscles \cite{fidalgo2021electromyographic,marco2017surface}. Besides, features from additional signals provide complementary information and are combined with EMG features. Taelman et al. \cite{taelman2011estimation} proposed to evaluate muscle fatigue by combining EMG features and tissue oxygenation features detected by near-infrared spectroscopy (NIRS). Studies like \cite{ce2017changes,anders2019inter} analyzed muscle fatigue through fusing EMG and mechanomyography (MMG). In this way, different aspects of muscle fibre activation can be included. Guo et al. \cite{guo2022assessment} developed a sensor system that integrated EMG, MMG and NIRS, and further evaluated fatigue-related features of each signal. Although inspiring, 
given the characteristics of daily exercise that involvess coordinated activation of multiple muscles, such features are limited by their capability of reflecting local muscle properties, hardly considering the nervous system adaptations of multi-muscle coordination. A more global feature integrating the neural control of multiple muscles would be beneficial.  Moreover,  such studies solely focus on the fatigue-related features and ignore the requirement of an estimator that fuses the features of multiple muscles into a score to continuously depict fatigue progression.

Except for fatigue-related features, an estimator is needed to fuse features and form a continuous fatigue score. Current methods mostly focus on static postures or isokinetic contractions of limited mucles. Monir et al. \cite{moniri2020real} utilized CNN to forecast EMG features of fatigued trunk muscles with the potential of estimating the binary fatigue states.  Guo et al. \cite{guo2021weak} proposed a  muscle fatigue estimator based on the weak monotonicity of features, which was able to form a continuous fatigue score estimator. Similarly, Yang et al. \cite{jiang2021data} trained a CNN to classify the onset of fatigue using signals of force plates and IMUs.  Xu et al. \cite{xu2018fatigue} empirically estimated a fatigue index of isometric contraction and used it to modify force estimation. Rocha et al. \cite{rocha2017weighted} proposed an estimator based on the assumptions of the Markov chain and the stationary process. It estimated   fatigue of isometric and isokinetic contractions by the deviation between the normalized features accumulation and the expectancy accumulation of a stationary process. Based on the same assumptions, Nascimento et al. \cite{de2022scalable} extended the framework and incorporated more features.  Other than just focusing on static postures or isokinetic contractions, such methods are limited by the assumption that the sEMG features of non-fatigued muscles are stationary, which might violate the non-stationary nature of sEMG during movements \cite{FARINA2000337}. The ability of accommodating daily exercise that involve dynamic and sub-maximal multi-muscle coordination should be further considered in designing the fatigue estimator. 

In this study, we consider the comprehensive characteristics of muscle fatigue during multiple muscle-involved daily movements and use walking as the test bed.  We propose to utilize features of muscle synergy fraction and spinal module activation in order to represent the temporal and spatial changes of multi-muscle activation caused by the fatigue-induced muscle compensation and spike timing deviations. We further synthesize the physiology-inspired features into a continuous fatigue score, in the manner of mathematically formulating the time-evolved nature and physiological characteristics of muscle fatigue as the loss function and the algorithmic architecture. To summarize, our contributions are as follows.
\begin{itemize}
	\item To the best of our knowledge, this is the first study that can continuously assess muscle fatigue for the daily exercise scenarios involving submaximal and dynamic contractions of multiple muscles. 
	\item We design physiology-inspired features to represent the fatigue-induced muscle compensation and spike timing deviations, in order to extract the global information of multi-muscle coordination.
	\item We develop  a physiologically rationale model  and formulate  the time-evolving dynamics of muscle fatigue as a novel loss function that solves the issue of lacking appropriate labels.
	\item We adapt the metrics of \cite{lu2020evaluating} to quantitatively evaluate our estimated fatigue and demonstrate it with extensive experiments.
\end{itemize}

\begin{figure*}[th]
	\centering
	\includegraphics[width = \textwidth]{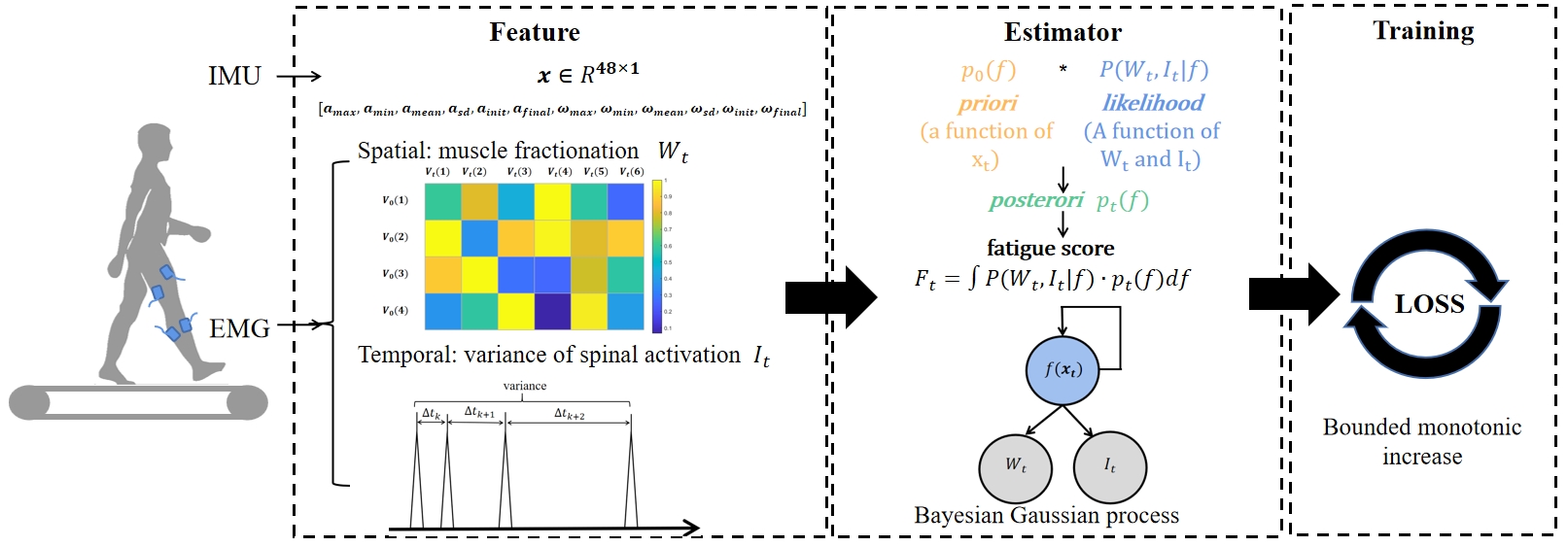}
	\caption{The schematic plot of the workflow. The exercise variable $\bm{x_t}$ is extracted from IMU measurements of shank and thigh. The fatigue-induced spatial and temporal patterns of multiple muscles $W_t$ and $I_t$ are extracted from EMG signals.Then, a directed graph model is built to infer the latent fatigue given the observations, Then, a physiology-inspired loss function is used to train the algorithm.}
	\label{workflow}
\end{figure*}

\section{Related Work}
We aim to develop a continuous muscle fatigue estimator for daily exercise that involves multi-muscle contraction. The estimator utilizes physiological signals from multiple muscles, which are sensed non-invasively. In the following, we review the potential challenges of proposing the method and the experience we learn from the literature.

Extracting \textbf{fatigue-related features} from physiological signals refers to utilizing the physiological knowledge of muscle fatigue and then forming the appropriate feature extraction methods.  Muscle fatigue is a multi-perspective concept and a whole body-involved process, which relates to the central command increase of motor regions, motoneuron spectral tuning, muscle fiber switch and motor performance alternation. On multi-muscle contraction, we can utilize, i.e. 1) the supraspinal/spinal neuronal changes of temporal activation patterns and  2) the muscle compensation phenomenon.  Specifically, repetitive activation of motoneurons and repetitive contractions of muscle fibers will induce reduced excitability of motoneuron itself and a supraspinal inhibition of motoneuron pool strengthened by group III/IV afferents and the short-latency reflex \cite{taylor2016neural}. In addition, despite the increased motor command sent by motor cortex partially compensates for the fatigue-induced muscle fiber deactivation, the supraspinal fatigue inhibits the further recruitment of muscle fibers and alter temporal activation patterns of the motoneuron pool \cite{taylor2008comparison, thomas2015central}, especially for low-intensity, long-duration, sub-maximal and whole-body exercise.  Moreover, for multi-muscle coordinated exercise, each muscle to different extents contributes to a given type of exercise, thus the muscle compensation will be induced by fatigue \cite{edouard2018sprint}. This corresponds to an altered spatial pattern of  multi-muscle activation.  Thus, it can be  beneficial to extract features inspired by the temporal and spatial changes of the multi-muscle activation to represent the fatigue of daily exercise.

Formulating an \textbf{estimator for muscle fatigue} requires an inverse model  that  uses the features of physiological signals to estimate the time-evolved dynamics of fatigue during prolonged exercise. Previous studies modelled the general principle of the muscle fatigue and/or recovery (expressed in terms of fatigue in pH in some studies) without using physiological sensors as input \cite{jebelli2020physiology, FREYLAW2021103104, frey2019optimisation}. Such models usually used the task load and/or working time as input and fitted the group-averaged endurance time. Although promising, the method is limited by the group-averaged output. Compared with such models, the model we aim to develop is to depict how the muscle fatigue progressed during prolonged daily exercise and how the features from physiological signals can be used to infer the progression of muscle fatigue.  Alternative studies proposed data-driven methods that either provide binary fatigue classification  \cite{jiang2021data, moniri2020real} or measure the deviation of EMG distributions between a relatively stationary process and the cumulative curve of EMG features \cite{rocha2017weighted,guo2021weak}. The latter approach that utilizes the stationary process violating the non-stationary nature of EMG during muscle contractions. As our comparison below, such methods do not present a satisfactory performance on evaluating the muscle fatigue on walking. It is necessary to model the dynamic progression of muscle fatigue during prolonged exercise, relate the model with the empirical observations  of fatigue-related features and then reverse the model to estimate the model parameters by the empirical observations.

\textbf{How to measure muscle fatigue} relates to how to get the label to train our algorithm and what metrics we can use to evaluate the performance of our algorithm.  Fatigue is a hidden state of human motor system, which is challenging to measure \cite{marco2017surface}. It should be noted that the fatigue induced by prolonged exercise relates to several levels of the physiological system. The evoked potential-based methods can measure either cortical or muscle fibre fatigue using stimulation \cite{gruet2013stimulation}. When the simulation is performed on motor cortex, the evoked potential of EMG is to measure how fatigue affects the information transfer inside the neural system. Alternatively, when the stimulation is performed on a muscle, the evoked potential of EMG is to measure how fatigue affects the contraction characteristics of the muscle fibres. Moreover, ATP metabolism and oxidative stress-based methods can measure local biochemical changes \cite{siegel2008cardiac}.  However, the abovementioned methods can only monitor the fatigue of either one muscle or the single muscle-involved corticomuscular system, which do not meet our requirement of assessing "integrative" fatigue for multiple muscles during daily exercise. 

Assessing multi-muscle fatigue by the decline of motor performance or by subjective feelings can be alternative approaches, that can form a score for multiple muscle-related exercise \cite{mugnosso2018muscle,sanchez2011velocity}. Detecting the decline of motor performance requires to use the same motor task for assessing fatigue as that for inducing fatigue, given that fatigue presents a task-specific output  \cite{granacher2010effects}. For our task, the motor performance of walking include over 12 gait parameters \cite{dos2020effects}, which are difficult to measure the changes of each parameter and then form a fatigue score. For subjective feelings, the endurance time of walking or running involves not only muscle fatigue, but also subjective rates of efforts and will power \cite{radel2017saving, hureau2018sensory}. But the subjective feelings obtained by self-reported questions although can summarize fatigue as some unified Likert-scale scores,   are insufficiently objective nor continuous. Lacking approximate approaches to provide the gold standard for the unified fatigue score of multi-muscle exercise relates to the challenges of training the algorithm in a supervised manner and evaluating the performance of the algorithm.
The methodology we develop to overcome the major challenges can be summarized as follows.
\begin{itemize}
	\item For \textit{fatigue-related features}, we utilize muscle synergy to represent the across-muscle coordination \cite{cheung2020plasticity}  and extract features from muscle synergy to represent the fatigue-induced muscle compensation.  In order to represent the altered temporal activation patterns of the motoneuron pool from bipolar EMG sensors, we can extract features of the spinal module activation, shown in our previous study \cite{yi2021bipolar}. 
	\item For \textit{muscle fatigue estimator},  we propose a Bayesian network to model the dynamics of muscle fatigue during prolonged exercise. We treat the fatigue-related features as the observations and estimate fatigue score by recursively applying the Bayesian  rule.  
	\item In this study,  we utilize a basic rule to form the loss function and then to \textit{train the algorithm} in a semi-supervised manner. That is, the fatigue score during prolonged walking should be not decreasing, i.e. weakly monotonically increasing. 
	\item We adopt the metrics to quantify the weak monotonicity of curves and then to \textit{evaluate our algorithm}. Moreover, although different views (e.g. decline of motor performance, changes of EMG signal distribution and subjective feelings) of fatigue depict different aspects, such different views follow similar trend during prolonged exercise \cite{weavil2019neuromuscular, hureau2018sensory}. Thus we also evaluate our algorithm by the similarity of the estimate  fatigue score evolving with time and the different views of fatigue.
\end{itemize}

\section{Method}

In this section, we  introduced how we score muscle fatigue of walking through formulating physiological principles of muscle fatigue into a computational model.  The following sessions introduce how we utilize the physiological principles of the time-evolving fatigue during prolonged walking to extract features, develop fatigue estimator, and evaluate through camparing with other views of fatigue.

\subsection{Data Collection}
\label{collection}
\begin{figure}[htbp]
	\centering
	\includegraphics[width = 9cm]{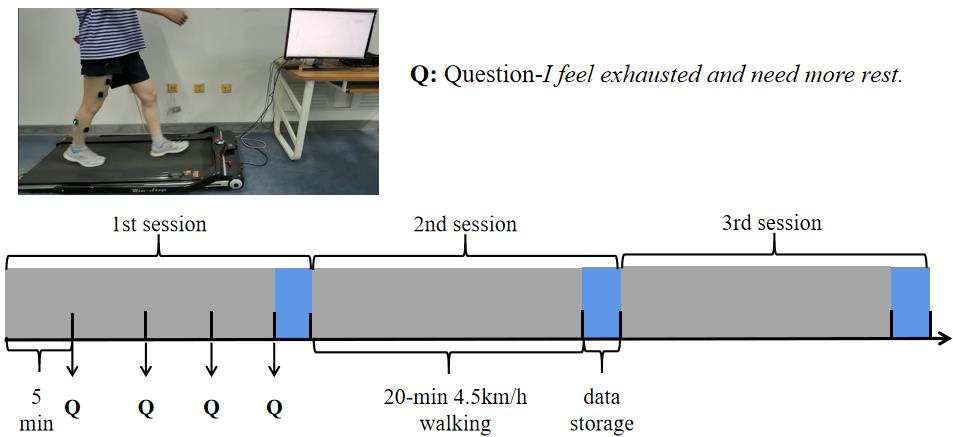}
	\caption{The schematic plot of the experimental paradigm. The whole experiment consists of three sessions, each lasting 20 minutes (dark grey) and followed by data storage periods (blue). For each session, the subjects are asked to walk with the speed of 4.5 km/h and  the question of  fatigue feelings is asked every 5 minutes. }
	\label{experiment}
\end{figure}

Ten subjects (5 males and 5 females, age range: 18-45 years old, weight range: 45-100 kg) were recruited and asked to walk at the speed of 4.5 km/h on a treadmill for three sessions. As shown in Fig. \ref{experiment}, each session consists of a 20-minute walking. Every 5 minutes, the subjects were asked to report their physical fatigue feelings using a 7-point Likert scale in response to the question: I feel exhausted and need more rest. The left-most point corresponded to “Strongly Agree” and the right-most point corresponded to “Strongly Disagree”. In order to exclude potential confounding factors \cite{muller2021neural}, we set no rewards nor punishment for fulfilling the task. Sitting down was not allowed between sessions, and the time between sessions was for the data storage purpose and lasted for maximum 2 minutes. We conducted the same experiment at a separate day. Two of the ten subjects came back 2 days later. Three of the ten subjects came back 3 days later. One subject came back 5 days later. The rest of them came back a week later. We do not constrain the same placement of EMG sensors on separate days. Each subject signed the informed consent before the experiment and the experimental protocol was approved by Chinese Ethics Committee of Registering Clinical Trials (ChiECRCT20200319).

Bipolar EMG sensors (Trigno Wireless System; DELSYS, Boston, MA, USA) were placed on the surface of target muscles after skin preparation through palpation. The 9 muscles are rectus femoris (RF), vastus lateralis (VL),vastus medialis (VM), tibialis anterior(TA), soleus (SOL), semitendinosus (ST), biceps femoris (long head ,BF), gastrocnemius lateralis (LG), gastrocnemius medialis (MG).  The sampling rate of EMG signals was 1111.11 Hz. Foot pressure sensors were attached to the heel and first metatarsal bone of the subject for phase labeling (swing phase, initial contact, midstance and propulsion), with the sampling rate of 500 Hz. The two signals were synchronized by a trigger device.

\subsection{Problem Formulation}

As stated in related work, the progression of muscle fatigue was time-evolved and related to the task type and training intensity ($P_1$). From the perspective of observation, fatigue induced the activation weight change of multiple muscles ($P_2$), i.e. muscle compensation,  and the supra-spinal inhibition of motor command and a group-level increase of the variance of motoneuron spike timings ($P_3$). From the perspective of fatigue dynamics, fatigue presented a weak increasing monotonicity ($P_4$). That is, the fatigue score $F_t$ generally increased with the increase of training time, but tolerated small jitters in the neighbourhood. And the increase of fatigue in the neighbourhood should be bounded.  Inspired by such prior, the problem can be formulated as:
\begin{align}
	\label{formulation}
	& \quad \quad F_t = g(W_t, I_t, \bm{x_t},F_{t-1})\\
	&\begin{array}{r@{\quad}r@{}l@{\quad}l}
		\nonumber s.t.   F_t =  F_{t-1} + \delta_{t-1}, \vert \delta_{t-1} \vert  \leq    \Delta
	\end{array}
\end{align}
where $g$ denoted a fatigue scoring function, $t$ denoted the time instant, $\bm{x_t}$ denoted the exercise variable corresponding to the prior of the task type and training intensity $P_1$,  $W_t$ denoted the activation weight change of multiple muscles corresponding to the muscle compensation prior $P_2$,   $I_t$  denoted the spike timing-related changes corresponding to the prior $P_3$, $\delta_{t-1}$ and $\Delta$ denoted the jitter tolerance and the bounded increase corresponding to the prior of bounded increase of fatigue in the neighbourhood $P_4$.

\subsection{Features Extraction}

In this part, we addressed the issue of the model input, i.e. the features corresponding to compensation ($P_2$) and group-level activation of spinal neurons ($P_3$) involved in walking-like multi-muscle coordination.  To do so, we factorized the multi-muscle signals into muscle synergies to capture muscle compensation and temporal patterns to capture the spinal module activation \cite{yi2021bipolar}. Spinal module activation reflects the group-level co-activation of spinal inter-neurons, thus can be used to extract the spike timing adaptations of fatigue. We followed the experience of literature and extract muscle synergies \cite{chvatal2013common,kieliba2018muscle} from EMG signals to represent multi-muscle coordination. The measured EMG signals were first band-pass filtered by a zero-phase 6th order Butterworth with cut-off frequencies of 20 and 500 Hz, then rectified to obtain the envelope, then low-pass filtered by a zero-phase 4th order Butterworth with a cut-off frequency of 10Hz and finally used to calculate muscle activation by a recursive filter and an exponential shaping function \cite{potvin1996mechanically}. Then, we used the non-negative matrix factorization (NMF) to factorize the muscle activation matrix M ($m \times t$,  $m$ was the number of muscles and $t$ was the number of samples). 
\begin{align}
	\label{NMF}
	M = V \cdot C + e
\end{align}
where $V$ ($m \times n$,  $n$ was the number of muscle synergies, also the number of spinal modules) was the muscle weighting components, $C$ ($n \times t$) was the temporal pattern components and $e$ was the residual error matrix. The weight matrix $W$ denoted how multiple muscles coordinate with each other in a spatial manner, each column of which denoted a spatial pattern of muscles corresponded to a muscle synergy.  Herein, we followed the experience of \cite{cheung2020plasticity} and used the muscle synergy fractionation $Frac_t$   to depict the muscle compensation induced by fatigue at each time instant $t$. 
\begin{align}
	\label{frac}
	Frac_t(i,k) = \bm{V_0}(i)^T\cdot \bm{V_t}(k)
\end{align}
where $\bm{V_0}(i)$ denoted the $i$th column of the initial weight matrix $V_0$, $\bm{V_t}(k)$ denoted the $k$th column of the latest factorized weight matrix $V_t$. We used the similarity between the muscle synergies of the initial sliding window (i.e. original muscle coordination manner) and those of the latest sliding window (i.e. the muscle coordination manner manipulated by fatigue) in order to estimate the merging and fractionation of multi-muscle spatial patterns. Then, we calculated the feature of muscle compensation, $W_t$, as the Frobenius norm of $Frac_t$. In this way, the muscle compensation of fatigue can be depicted.

 Treating the time-varying coefficient matrix $C$ as the smoothed version of spinal module activation \cite{yi2021bipolar}, we extracted the supra-spinal inhibition and the increased group-level variation of spike timings from each row of $C$. Specifically, as presented in our previous work, we first binarized each row of $C$ by thresholding its amplitude. The threshold was calculated as the average plus one standard deviation. Each binarized sequence denoted the co-activation of the functionally grouped interneurons, i.e. the activation of each spinal module. Then, we pooled the binarized sequences into one sequence to denote the activation train of spinal modules. We used $1$ as the spiking of spinal modules and $0$ as de-activation of spinal modules. Finally, we calculated the standard deviation of the interval between each spinal module's spiking. That is, the standard deviation of the number of $0$ between two nearest $1$ was calculated. In this way, the spinal and group-level activation changes reflected by the variation of spinal module spike timings were depicted. We denoted the feature by $I_t$.

\subsection{The Fatigue Estimator}
\begin{figure}[th]
	\centering
	\includegraphics[width = 3.5cm]{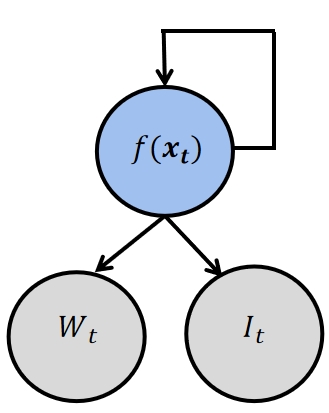}
	\caption{The schematic plot of fatigue dynamic model, corresponding to how the hidden state muscle fatigue $f_t$ induces the muscle compensation $W_t$ and the variation of spinal module spike timings $I_t$.}
	\label{dynamic}
\end{figure}

According to the physiology-inspired problem formulation, the phenomenological view enabled us to capture the natural dynamics of fatigue and to model it as a Markov progress \cite{bishop2006pattern, murphy2002dynamic}. As shown in Fig. \ref{dynamic}, we further treated it as a time-evolving directed graph model where the fatigue state $F_t$ was a time-evolved hidden state and affected the  observational states of muscle compensation $W_t$ and the variation of spinal module spike timings $I_t$.

The function $g$ in Eq. (\ref{formulation}) served to update the posterior of fatigue scores given the observations of muscle compensation and spinal module activation variation. To form the posterior updating paradigm, we introduced a latent state $f$ and estimate the muscle fatigue score as follow.
\begin{align}
	\label{frac}
	F_t = \int P(W_t,I_t|f) \cdot p_t(f) df
\end{align}
where $p_t(f)$ was the posterior of the latent state $f$ at the time instant $t$.  It can be obtained recursively in a Bayesian manner, given the exercise variable $\bm{x_t}$ and the observations of muscles $W_t$ and $I_t$.
\begin{align}
	\label{posterior}
	p_t(f) = \frac{P(W_t,I_t|f,\bm{x_{t-1}}) \cdot p_{t-1}(f|\bm{x_{t-1}})}{\int P(W_t,I_t|f,\bm{x_{t-1}}) \cdot p_{t-1}(f|\bm{x_{t-1}}) df}
\end{align}
where $W_t$ and $I_t$ were independent given $f$. Note that for every term with lower letter $t$ was calculated in non-overlapped sliding windows with the same length. The length of the sliding window will be campared and selected in Section \ref{window length}.

\subsubsection{The Prior Capturing The Task Type And Training Intensity}

We placed a Gaussian prior over the latent state $f$.
\begin{align}
	\label{p0}
	p_0(f) = N(m(\bm{x}),k(\bm{x},\bm{x}'))
\end{align}
where $m(\bm{x})$ denoted  the weighted mean of the elements of $\bm{x}$, $k(\cdot ,\cdot )$ denoted the covariance function. We denoted $\bm{x_t}$ by the IMU features commonly used for classifying gait phases and locomotion modes. The IMU features were cascaded by the traditionally used features extracted from the angular rates and accelerations \cite{barshan2014recognizing} measured by the two IMUs mounted on shank and thigh, respectively. The dimension of a feature was 48. Herein, we used a linear mean function, $m = \bm{\beta}^T \bm{x_t}$, where $\bm{\beta}$ was the learnable parameter. And we used the squared exponential kernel to project the exercise variable into a low-dimensional manifold. 
\begin{align}
	\label{kernel}
	k(\bm{x},\bm{x}')) = exp(-(\bm{x}-\bm{x}')^T D(\bm{x}-\bm{x}'))
\end{align}
where $D$ denoted the diagonal matrix ($48 \times 48$) whose diagonal elements $d_{i}$ were the variance of $(\bm{x}-\bm{x}')$. 

\subsubsection{The Likelihood of Observation}

In order to form a probit likelihood, we used the cumulative density function of a standard normal $\Phi(y) = \int_{-\infty}^{y}exp(\frac{-x^2}{2}) dx $. Specifically, the observation likelihood can be formulated as 
\begin{align}
	\label{likelihood}
	P(W_t,I_t|f,\bm{x_t}) = \frac{sgm(W_t)\cdot sgm(I_t) \cdot (f-m(\bm(x_t)))}{\sigma_n^2}
\end{align}
where $sgm(\cdot)$ denoted the sigmoid function that squashed the variable inside the bracket into $[0,1]$. Other likelihood functions that presented a probit form can also be used.

\subsubsection{Posterior Updating Via Gaussian Process Approximation}

Directly updating the posterior of the latent function by Eq. \ref{likelihood} was non-Gaussian and hard to have a closed-form solution.  In order to iteratively update the posterior, we approximated the posterior $p_t(f)$ by the Gaussian distribution that has the smallest Kullback–Leibler (KL) divergence with it. We used the Laplace approximation technique to match the first two moment of the two distributions and can obtain the following update equations.
\begin{align}
	\label{up1}
	\mu_t(\bm{x_t}) &= \bm{\alpha_t^T}\bm{k(x_t)}
\end{align}
\begin{align}
	\label{up1-1}
	k_t(\bm{x_t},\bm{x}') &= k(\bm{x_t},\bm{x_t}') + \bm{k(x_t)}^T C_t \bm{k(x')}
\end{align}
where $\bm{k(x_t)}=[k(\bm{x_{t-T}},\bm{x_t}),..., k(\bm{x_{t-1}},\bm{x_t})]$, $\bm{x}' = \frac{1}{T} \sum_{i=1}^T \bm{x_{t-i}}$, $\bm{\alpha_t}$ (a $T$-dimensional vector) and $C_t$ ($T \times T$) can be calculated by
\begin{align}
	\label{up2}
	\bm{\alpha_t} &= \bm{\alpha_{t-1}} + \gamma_1 C_{t-1}\bm{k_t}
\end{align}
\begin{align}
	\label{up2-1}
	C_t &= C_{t-1} + \gamma_2 (C_{t-1}\bm{k_t})(C_{t-1}\bm{k_t})^T
\end{align}
where $\bm{k_t} = [k(\bm{x_{t-T}},\bm{x_{t-1}}),..., k(\bm{x_{t-1}},\bm{x_{t-1}})]$, $\gamma_1$ and $\gamma_2$ can be calculated by 
\begin{align}
	\label{up3}
	\nonumber \gamma_1 &= \partial log \int P(W_t,I_t|f,\bm{x_t})df / \partial f \\
	\gamma_2 &= \partial^2 log \int P(W_t,I_t|f,\bm{x_t})df / \partial f^2 
\end{align}

Herein, we set $T$ as 50.

\subsubsection{ Fatigue Scoring }
Given $\bm{\alpha_t}$ and $C_t$, we can obtain the close-form solutions of the posterior at each time instant, thus can have the fatigue score. 
\begin{align}
	\label{score}
	\nonumber F_t &= \int P(W_t,I_t|f) \cdot p_t(f) df \\
	&= \Phi(\frac{sgm(W_t)\cdot sgm(I_t)\cdot (\mu_t-m(\bm{x_t}))}{\sigma_x^2})
\end{align}
where $\sigma_x = \sqrt{\sigma_n^2 + k(\bm{x_t},\bm{x}')}$. In this way, we formed the fatigue estimator using Bayesian Gaussian Process and can estimate the time-evolving muscle fatigue at each time instant.

\subsection{Training}
As mentioned in related work, the supervision information for training the model can hardly be obtained. It was reported in \cite{weavil2019neuromuscular} that the subjective fatigue feelings and metabolite changed like blood lactic acid can either reflect the global inner feeling or the biochemical body fluid changes of local muscles. Neither of them was a biased measurement of nervous system adaptation induced by muscle fatigue, especially for multi-muscle coordinated exercises, and can not be treated equally. Moreover, both the subjective feelings and the metabolite changes were measured discretely in each time section. That is, neither the time resolution nor the measurement method of them can contribute to a continuous measurement. Thus, we turned to formulating the loss function using the principles of the fatigue itself. The development of fatigue during exercise follows the following rules.
\begin{itemize}
	\item Fatigue increases as the exercise goes on. 
	\item The muscle fatigue should increase gradually. That is, the increase in the neighborhood should be bounded. 
\end{itemize}
The loss function can thus be formulated as 
\begin{align}
	\label{loss}
	L = \sum_p (\Delta (F_{p_1},F_{p_2}) + (1-F_{p_1})^2 + (1-F_{p_2})^2 )
\end{align}
\begin{align}
	\nonumber \Delta (F_{p_1},F_{p_2}) = ReLu((&(F_{p_1}-F_{p_2}-(p_1-p_2)\cdot \delta)^2)\\
	\nonumber &-((p_1-p_2)\cdot \delta)^2)
\end{align}
where $p$ denoted the $p$th pair of time instants $(p_1,p_2),p_1>p_2$, $F_{p_i}, i=1,2$ denoted the fatigue score at the $p_i$th time instant, $\Delta (F_{p_1},F_{p_2})$ denoted the weak monotonicity that tolerates jitters if $p_1$ is close to $p_2$. 

\begin{figure}[th]
	\centering
	\includegraphics[width = 3.5cm]{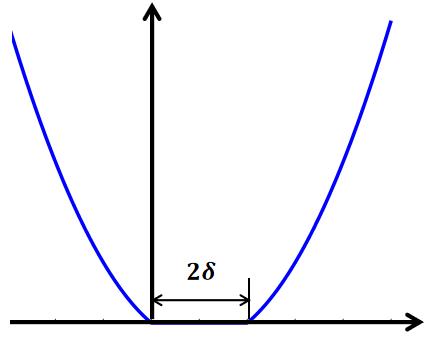}
	\caption{The schematic plot of $y=\Delta(x)$.}
	\label{fig:loss}
\end{figure}
As shown in Fig. \ref{fig:loss}, $\Delta(x) = ReLu((x-\delta)^2-\delta^2)$ can tolerate the increase bounded by $2\delta$. In Eq. (\ref{loss}), the bound was further scaled by the time distance of the index pair $(p_1,p_2)$.  With the loss function, we trained the learnable parameters $\sigma_n$, $\bm{\beta}$ and $D$ that were shared among subjects in an end-to-end manner. Specifically, we generated data pairs corresponded to the index pair $(p_1, p_2)$ in the training set and the validation set.  The Adam optimizer was used. And the learning rate was set as 0.1. The algorithm was trained with the Adam optimizer, the learning rate of 0.1 and the batch size of 100 pairs. The training stopped when the decrease of loss between epochs was lower than 0.001 for over 3 times or the maximum 1000 epochs were reached or early stopping criteria was reached on the validation set. We split the whole data set into training, validation and test set following a 7:1:2 strategy. For each epoch, $\bm{\alpha_t}$ and $C_t$ were initialized as a vector whose elements were randomly set in the range of $(0,0.2)$ and a diagonal matrix whose diagonal elements were randomly set in the range of $(0,2)$. During the testing phase, we randomly initialized $\bm{\alpha_t}$ and $C_t$ using the same strategy 100 times in order to average the effects of the initialization. All the data were processed by MATLAB 2015b. The settings of the laptop we used were Intel (R) Core (TM) i7-10510U CPU $@$ 1.80GHz   2.30 GHz, RAM 8.00 GB (Dell Inspiration 5598). The average training time for the proposed Bayesian network was 58.112 seconds.

\begin{breakablealgorithm}
	\caption{Fatigue Score Estimation}
	\begin{algorithmic}
		\renewcommand{\algorithmicrequire}{\textbf{Input:}}
		\renewcommand{\algorithmicensure}{\textbf{Output:}}
		\State $\bm{Input:} D, \bm{\beta}, \sigma_n$, IMU and EMG signals
		\State $\bm{Output:} F_t$ 
		\For {each time window $t$}
		\State  IMU and EMG signals inside window $\gets$ data windowing
		\State 
		\State \textit{Input preparation}
		\State  $M\gets$ Filtering the EMG signals and calculating the muscle activation
		\State  $\bold{x_t} \gets$ Feature extracted from IMU signals
		\State  $V, C\gets M$     \Comment{NMF,  Eq. (\ref{NMF})}
		\State $W_t \gets V$ \Comment{Eq. (\ref{frac}), and the Frobeius norm}
		\State $ I_t \gets C$ \Comment{Thresholding, binarizing and calculating the std of zeros}
		\State 
		\State  \textit{Calculating the interediate variables}
		\If{the first time window,  $t =0$} 
		\State{Initialize $\bm{\alpha_0}$ and $C_0$ 100 times randomly}
		\EndIf
		\State set $T = 50$
		\State $m(\bm{x_t}) \gets \bm{\beta}^T \cdot \bm{x_t}$
		\State $\bm{x'} \gets \frac{1}{T} \sum_{i=1}^T \bm{x_{t-i}}$
		\State $k(\bm{x_t},\bm{x'}) \gets \bm{x'} , \bm{x_t}$ \Comment{Eq.(\ref{kernel})}
		\State $\bm{k_t} \gets [k(\bm{x_{t-T}},\bm{x_{t-1}}),..., k(\bm{x_{t-1}},\bm{x_{t-1}})]$
		\State $\bm{k(x_t)} \gets [k(\bm{x_{t-T}},\bm{x_t}),..., k(\bm{x_{t-1}},\bm{x_t})]$
		\State $\gamma_1,  \gamma_2 \gets W_t,  I_t, \bm{x_t}$ \Comment{Eq.(\ref{up3})}
		\State $\bm{\alpha_t} \gets \bm{\alpha_{t-1}}, \gamma_1, C_{t-1}, \bm{k_t}$ \Comment{Eq.(\ref{up2})}
		\State $C_t \gets  \gamma_2,  C_{t-1},  \bm{k_t}$ \Comment{Eq.(\ref{up2-1})}
		\State $\mu_t \gets \bm{\alpha_t}, \bm{k(x_t)}$ \Comment{Eq.(\ref{up1})}
		\State 
		\State  \textit{Final output}
		\State $F_t \gets \sigma_n, k(\bm{x_t},\bm{x'}), W_t, I_t, \mu_t, m(\bm{x_t})$ \Comment{Eq.(\ref{score})}
		\State Average $F_t$ across all the random initializations
		\EndFor
	\end{algorithmic}
\end{breakablealgorithm}

\subsection{Evaluation}
As mentioned before, it is hard to have a gold standard to evaluate the "integrative" muscle fatigue estimation of multiple muscles during prolonged walking. Herein, we utilized the following facts to evaluate the performance of our estimated fatigue score: 1) muscle fatigue increases with weak monotonicity during prolonged walking; 2) the muscle fatigue estimated on different days but from the same subject should present robustness across days; 3) the muscle fatigue estimated on the same day but from different subjects should present robustness across subjects; 4) the fatigue depicted by different views, although presented different aspects of the physiological system related to fatigue, should present similar trend during prolonged walking \cite{weavil2019neuromuscular, hureau2018sensory}.

\subsubsection{Fatigue Depicted From Other Views}
In the following, we introduced the "intergrative" fatigue score of prolonged walking  depicted from other views, which may not be realtime but can be used for the performance evaluation. Specifically, we used the subjective feelings collected during the experiment (detiled in Section \ref{collection}) and the performance decay of gait phase classification.

For the subjective feeling, we down-sampled the estimated fatigue score to match the time instants of subjective feelings (SF). We quantified the trend similarity between SF and the down-sampled, estimated fatigue score for the performance evaluation. The metric used for the quanitification was presented in the following session. For the performance decay of gait phase classification, we extracted the classic feature set, Hudgin’s set \cite{englehart2003robust}, with the recommended window length of 128 ms and step of 15 ms. We trained the classifier with the first 2-minute data and tested its accuracy using the data in each 2-min, non-overlapped sliding windows of the following time. And we calculated the accuracy degradation (AD), i.e. to what extent the accuracy decreases in the following time compared with that of the first 2-min data.  Similarly, we down-sampled the estimated fatigue score and quantified the trend similarity between the down-sampled fatigue score and AD.

\subsubsection{Metrics} In the following, we adopted the metrics developed by \cite{lu2020evaluating, guo2021weak}.

\textbf{Weak Monotonicity (WO):}
For a continuous fatigue score $J_j$, we first calculated how many sample points of the score trajectory fall into the range of weakly monotonic increasing. The tolerance for the weak monotonicity was set as delta. The points that follow the weak monotonicity consist of $D_j^+$ and the other points consist of $D_j^-$.
\begin{align}
	\label{point}
	D_j^+ &= \{ J_j(t)|J_j(t)>J_j(t-1)+\delta_j(t-1) \} \\
	\nonumber D_j^- &= \{ J_j(t)|J_j(t) \notin D^+ \}
\end{align}
where $\delta_j$ should accommodate the variation of the trajectory $J_j$, thus can be calculated as the standard deviation of the trajectory, i.e. $\delta_j = std(J_j)$. 
We further calculated the number of points in $D_j^+$ and $D_j^+$ as $Num^+$ and $Num^-$, respectively. Then, we calculated WO by 
\begin{align}
	\label{WO}
	WO_j = \frac{Num^+}{N-1} - \frac{Num^-}{N-1}
\end{align}
where $N$ denotes the total number of the sample points of the trajectory.

\textbf{Trendability (Tr):} It is assumed that different views of fatigue, although present different perspectives, should follow some common trends. In addition, common trends should also exist among subjects and among different days. The trendability can be calculated as the similarity between two trajectories. We evaluated the trendability by
\begin{align}
	\label{Tr}
	Tr_{1,2} = Corr(J_1,J_2)
\end{align}
where $Corr(\cdot)$ denotes the Pearson coefficient, $J_1$ and $J_2$ denote the metrics measured  by different views (e.g. by the fatigue estimator, the accuracy degradation of EMG-based gait phase classifier and the fatigue feelings), the estimated fatigue scores in different days or the  estimated fatigue scores from different subjects.

\textbf{Suitability (S):} Suitability combines weak monotonicity and trendability to evaluate the fatigue score comprehensively.
\begin{align}
	\label{S}
	S = (WO_1+WO_2)\cdot Tr_{1,2}
\end{align}
where $WO_1$ and $WO_2$ denote the weak monotonicity measured  by different views (e.g. by the fatigue estimator, the accuracy degradation of EMG-based gait phase classifier and the fatigue feelings) or in different days.

\section{Results}
We took an initial step toward continuously assessing muscle fatigue under the scenario of  sub-maximal and dynamic contractions of multi-muscles. In this section, we evaluated the proposed method by testing its cross-day stability, similarities with other views of fatigue (i.e. performance decay of gait phase classification, subjective feelings) and similarities among subjects. We performed extensive experiments on selecting the sliding window length, input features, comparing with other algorithms on trendability with other views of fatigue and cross-day stability and cross-subject stability.

\subsection{Lengths of The Sliding Window}
\label{window length}
\renewcommand{\dblfloatpagefraction}{.9}
\begin{figure}[hbp]
	\centering 
	\subfigure[$WO$]{
		\includegraphics[width=4.0cm]{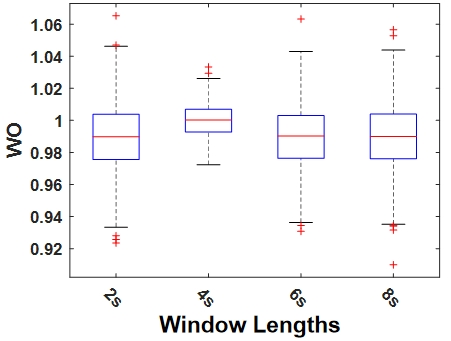}}
	\subfigure[$Tr_{AD,F}$]{
		\includegraphics[width=4.0cm]{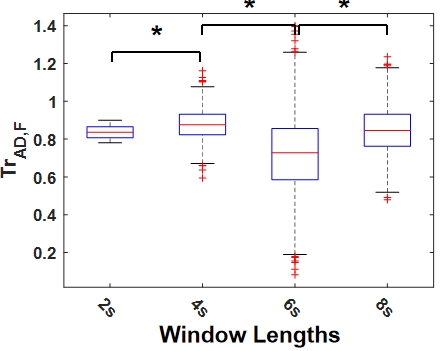}}
	\subfigure[$Tr_{SF,F}$]{
		\includegraphics[width=4.0cm]{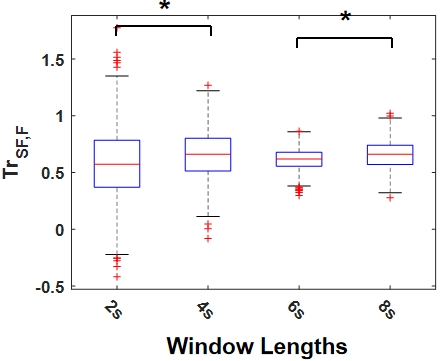}}
	\caption{The performance of different lengths of windows.  $*$ indicates a statistically significant difference with other conditions (paired t-test, $\alpha <$0.05). $WO$ denotes the weak monotonicity, $Tr_{SF,F}$ denotes the similarity between subjective fatigue feelings and the estimated fatigue scores, $Tr_{AD,F}$ denotes the similarity between accuracy degradation and the estimated fatigue scores.} 
	\label{fig:win}  
	\vspace{-1.5em}
\end{figure}
As we used the non-overlapped window for the fatigue estimation, we thus compared the performance of different window lengths without worrying about the   steps between windows. We calculated WO for the fatigue score estimated by different sliding window lengths. Given that the average duration of each step is 0.8s, the window lengths we set were approximately 2.5 steps for 2s,  5 steps for 4s, 7.5 steps for 6s and 10 steps for 8s. We performrf grid search and evaluate the performance by 1) calculating the $WO$s of each window length and 2) calculating the estimated fatigue score’s similarity with the subjective feelings and with the gait phase classification performance decay (i.e. $Tr_{AD,F}$ and $Tr_{SF,F}$). The metrics were averaged over subjects and days. Paired t-tests are used to test the significant difference. And the Shapiro-Wilk test is performed to test normality. Alpha level is set as 0.05.

As shown in Fig.\ref{fig:win}, WO does not present a significant difference among window lengths, while $Tr_{AD,F}$ and $Tr_{SF,F}$ present significant but small differences. It can be shown that the step of 4 seconds and 8 seconds present almost the same and better than other steps. Considering the time resolution of the method, we choose 4 seconds (about 5 steps of walking) as the final window length.

\subsection{Influence of Input Features}

\renewcommand{\dblfloatpagefraction}{.9}

We tested the influence of different features. Following the experience of using the root mean square (RMS) and the median frequency (MDF) to depict fatigue \cite{zhang2018non,wang2021novel}, we performed PCA on the RMS and MDF of the 9 muscles, respectively, and extracted the first principle components denoted by $P_{RMS}$ and $P_{MDF}$, respectively.  $P_{RMS}$ and $P_{MDF}$ were used as the inputs for the estimator to replace $W_t$ and $I_t$. We used $RM$ to denote using $P_{RMS}$ and $P_{MDF}$ as input features and used $SI$ to denote using $W_t$ and $I_t$ as input features. We presented the weak monotonicity comparison and use $Tr_{AD,F}$ and $Tr_{SF,F}$ versus $Tr_{AD,RM}$ and $Tr_{SF,RM}$ for the trendability comparison. Alpha level is set as 0.05.

\begin{figure}[h]
	\subfigure[$WO$]{
		\includegraphics[width=0.15\textwidth]{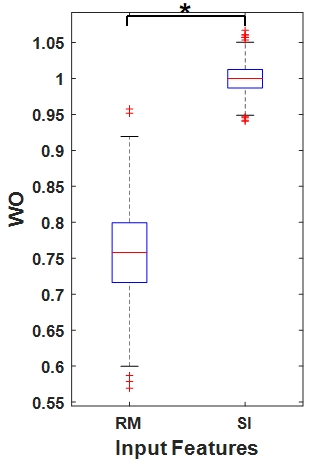}}
	\subfigure[$Tr_{AD,F}$]{
		\includegraphics[width=0.15\textwidth]{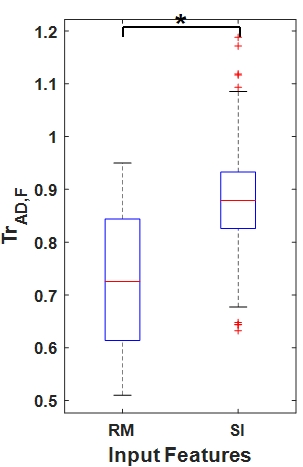}}
	\subfigure[$Tr_{SF,F}$]{
		\includegraphics[width=0.15\textwidth]{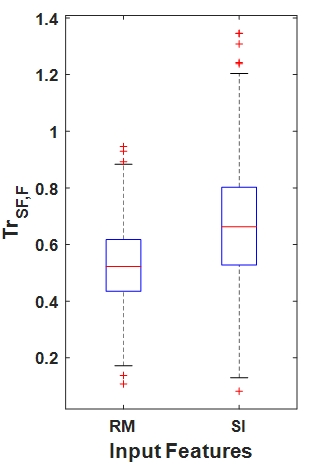}}
	\caption{The performance of different input features for our proposed estimator.  $*$ indicates a statistically significant difference (paired t-test, $\alpha <$0.05). RM denotes using the root mean square and the median frequency as input features, SI denotes using the muscle compensation and the spinal adaptation of spike timings as input features, $WO$ denotes the weak monotonicity, $Tr_{SF,F}$ denotes the similarity between subjective fatigue feelings and the estimated fatigue scores, $Tr_{AD,F}$ denotes the similarity between accuracy degradation and the estimated fatigue scores, $RM$ denotes using the first principle components of RMS and MDF as input features, and $SI$ denotes using $W_t$ and $I_t$ as input features.}
	\label{fig:feature} 
	\vspace{-1.5em}
\end{figure}

It can be seen from Fig. \ref{fig:feature} that our proposed features $W_t$ and $I_t$ generally outperform the features RMS and MDF traditionally used for estimating muscle fatigue under isometric or static contractions. The results indicate that solely depicting single muscle properties for multi-muscle contractions is insufficient.

\subsection{Fatigue Assessment}

We herein compared our proposed method with the methods that also use EMG signals to continuously assess muscle fatigue. We used the following methods for the performance comparison with our method. 1) We adopted the weighted-cumulative fatigue estimator (WC)  \cite{rocha2017weighted} that utilized the static and Markov assumptions of fatigue. Other than $W_t$ and $I_t$, we used the demonstrated input MDF and apply PCA on the muscles to obtain the first principle component. We denoted the weighted-cumulative fatigue estimator with different inputs as $WC_M$ for using MDF as input, $WC_W$ for using $W_t$ as input and $WC_I$ for using $I_t$ as input. 2) We used the weak monotonicity (WO)-based method to translate features into WO scores for depicting muscle fatigue \cite{guo2021weak}. We used the ratio of the points in a sliding window that do not follow the weak monotonicity principle as the metrics of fatigue. We also use the first principle component of MDF, $W_t$ and $I_t$ as the input features and denote them by  $WO_M$, $WO_W$ and $WO_I$. For a fair comparison purpose, the same windowing scheme is used. We performed the comparison by using the metrics of WO, $Tr_{AD,F}$ and $Tr_{SF,F}$ of each estimated fatigue score,  where   $Tr_{AD,F}$ denotes the trendability between the accuracy degradation and the estimated fatigue scores and  $Tr_{SF,F}$ denotes the trendability between the subjective fatigue feelings and the estimated fatigue scores.   Alpha level is set as 0.05.

\renewcommand{\dblfloatpagefraction}{.9}
\begin{figure}[htbp]
	\subfigure[$WO$]{
		\includegraphics[width=0.23\textwidth]{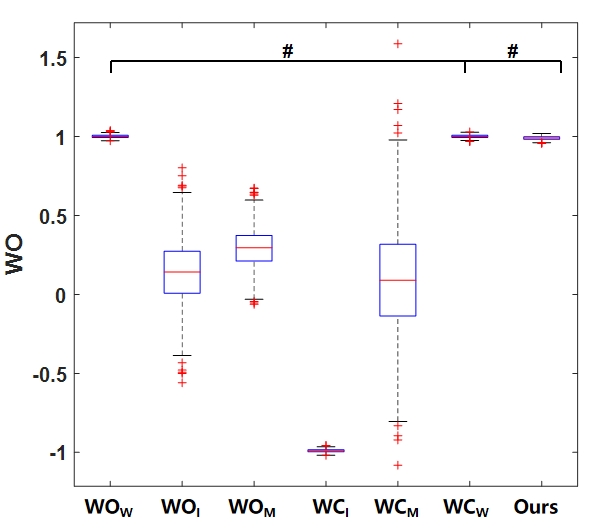}}
	\subfigure[$Tr_{AD,F}$]{
		\includegraphics[width=0.23\textwidth]{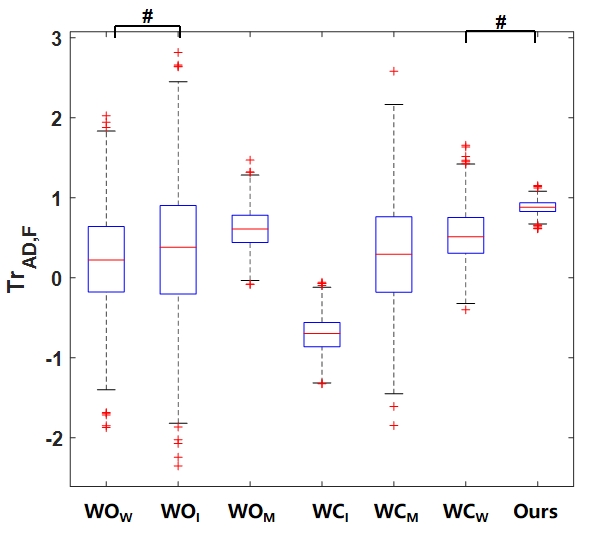}}
	\subfigure[$Tr_{SF,F}$]{
		\includegraphics[width=0.23\textwidth]{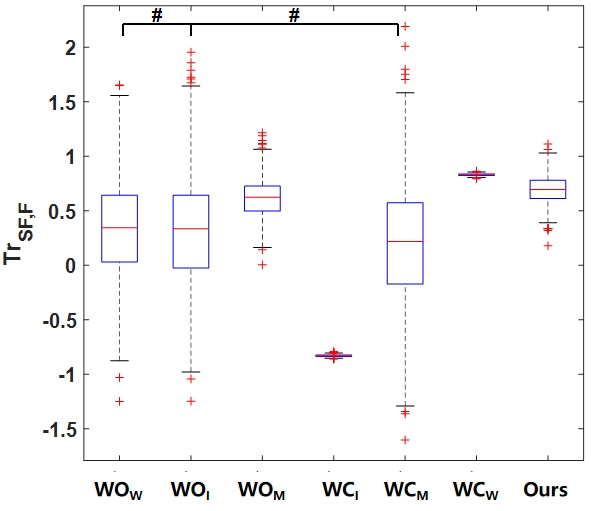}}
	\caption{The performance comparison of different methods.  $\#$ indicates there is no statistically significant difference (paired t-test, $\alpha \geq $ 0.05). $WO_W$, $WO_I$ and $WO_{M}$ denote the weak monotonicity-based  method with the input features of muscle compensation, the variation of spinal module spike timings and the median frequency, respectively. Similarly, $WC_{W}$, $WC_{I}$ and $WC_{M}$ denote the weighted cumulative fatigue estimator with the three input features. $Ours$ denotes the method proposed in this study.}  
	\label{fig:compare}  
	\vspace{-1.5em}
\end{figure}

As shown in Fig. \ref{fig:compare}, our method presents similarly good weak mononitivity with $WO_W$ and $WC_{W}$, but outperform them by $Tr_{AD,F}$ and the overall metric $S$. It can be shown in Fig. \ref{fig:compare} that although $WC_{W}$ presents a larger $Tr_{SF,F}$, there is no significance between $WC_{W}$ and our method. $WC_{I}$ presents negative WO and trendabilities, which indicates the potentially poor performance for continuously assessing muscle fatigue. Similarly, $WC_{I}$ presents a relatively large suitability, but suffers from small or even negative WO and Tr.  $WO_W$, $WO_I$, $WO_{M}$, $WC_{M}$ present relatively small values of trendabilities and suitabilities, which also indicate their worsened performance when applyed on the multi-muscle coordinated exercise.

\vspace{0.3cm}
\subsection{Cross-Day Stability}
We utilized the data collected from two separate days and estimated fatigue scores using different methods for each separate day. We tested the cross-day stability by comparing the trend similarity of each fatigue estimator between the two separate days, denoted by trendability ($Tr_{D1,D2}$). We also integrated the WO metric of each day's estimate and the trendability as suitability ($S_{D1,D2}$). We evaluated the cross-day performance of the above-mentioned methods.

\begin{figure}[t]
	\centering 
	\subfigure[$W_t$]{
		\includegraphics[width=0.15\textwidth]{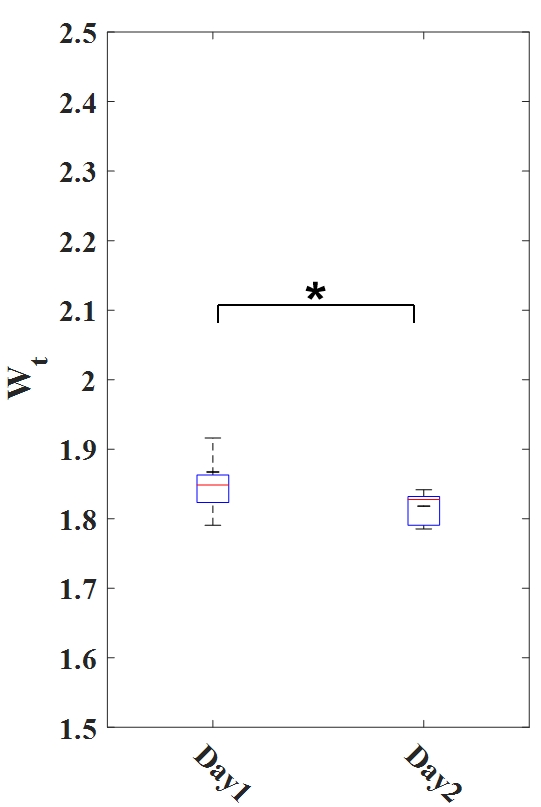}}
	\subfigure[$I_t$]{
		\includegraphics[width=0.15\textwidth]{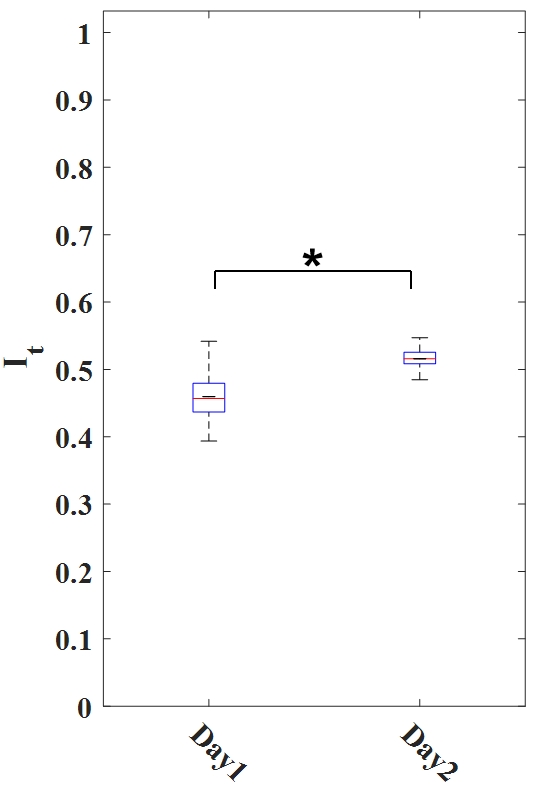}}
	\caption{The muscle difference across different days.  $*$ indicates a statistically significant difference (paired t-test, $\alpha <$0.05).} 
	\label{fig:mus_day}  
\end{figure}

\begin{figure}[t]
	\centering 
	\subfigure[$Tr_{D1,D2}$]{
		\includegraphics[width=0.23\textwidth]{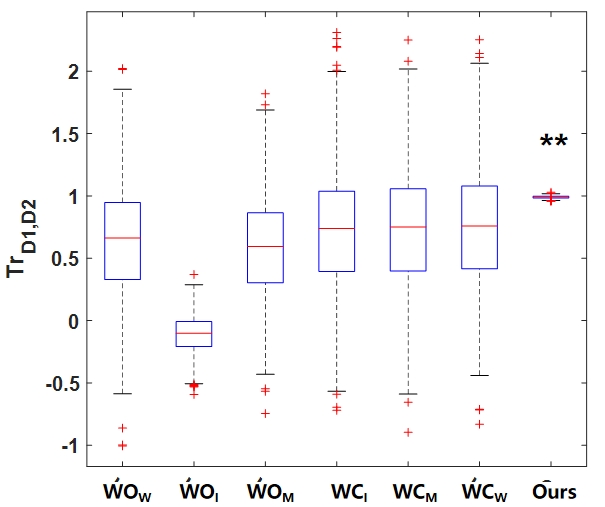}}
	\subfigure[$S_{D1,D2}$]{
		\includegraphics[width=0.23\textwidth]{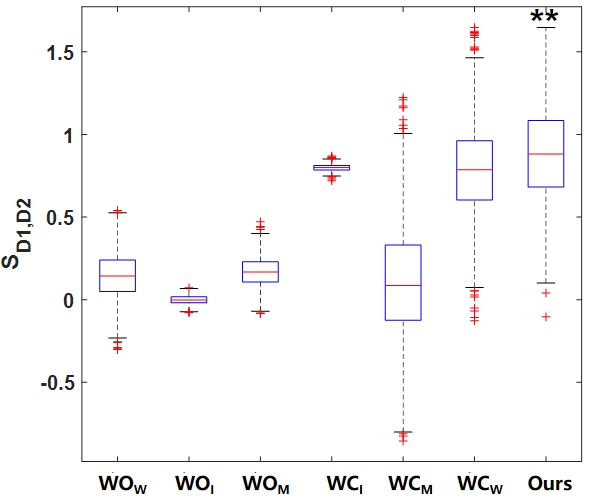}}
	\caption{The performance of different days.  $**$ indicates a statistically significant difference with other conditions (paired t-test, $\alpha <$0.05).} 
	\label{fig:day}  
	\vspace{-2em}
\end{figure}

As shown in Fig. \ref{fig:mus_day}, the features of muscle cmpensation and spinal module activation are different across days. As shown in Fig. \ref{fig:day}, the weighted cumulative estimator-based methods ($WC_{I}$, $WC_{M}$, $WC_{W}$) present a relatively stable performance under the cross-day scenario. This meets the performance demonstrated in \cite{rocha2017weighted} under isometric contractions. For the sub-maximal and dynamic contractions of multiple muscles, our method outperforms the weighted cumulative estimator-based methods by the trendabilities and suitabilities between days. In addition, the weak monotonicity-based methods ($WO_W$, $WO_I$, $WO_{M}$) present worsened performance, which indicates their insufficient stability for cross-day measurements.

\subsection{Cross-Subject Stability}
Herein, we evaluated the cross-subject stability for each fatigue estimator by calculating the trendability within each pair of subjects.

\begin{figure}[h]
	\centering
	\includegraphics[width =0.25\textwidth]{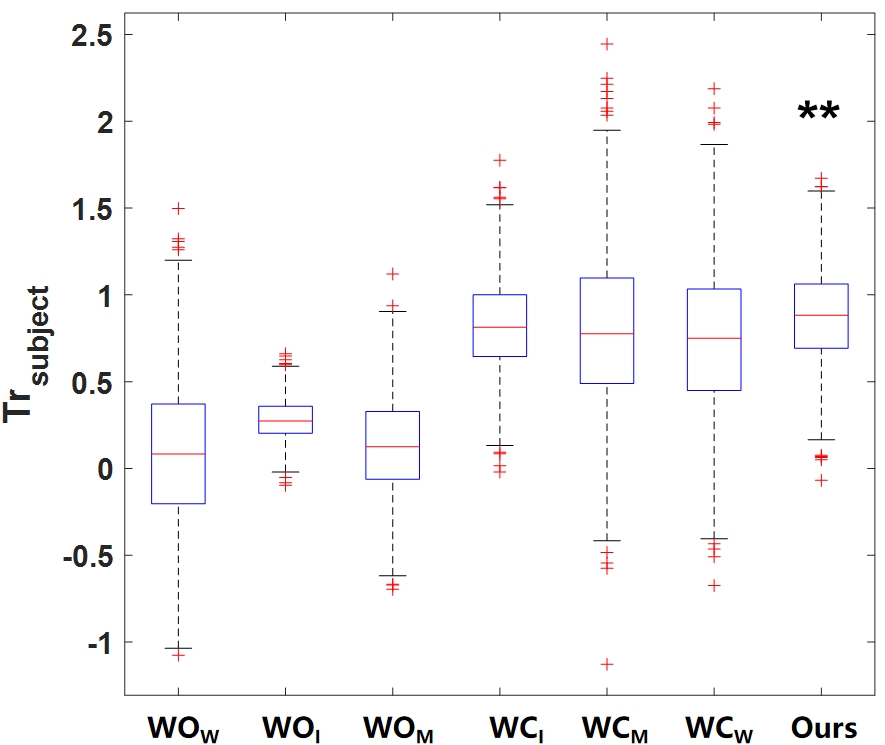}
	\caption{The trendability within each pair of subjects. $**$ indicates a statistically significant difference with other conditions (paired t-test, $\alpha <$0.05).}
	\label{fig:sub}
	\vspace{-1.5em}
\end{figure}

As shown in Fig. \ref{fig:sub}, our method significantly outperforms other methods by the cross-subject similarities. The progression of fatigue share some common trends among subjects, despite individual characteristics. For example, at least a monotonic increase of the estimated fatigue should be shared among subjects.

\section{Discussion}
In this study, we propose to depict the continuous progression of muscle fatigue under multi-muscle coordinated exercise and formulate the issue into a physiologically rationale model. We then solve the issue of lacking metrics to evaluate the fatigue estimator’s performance by adopting the metrics of weak monotonicity, trendability and suitability. Through extensive experiments, we evaluate our method’s performance of different hyper-parameters, different input features, cross-day stability and cross-subject stability. The results demonstrate our method’s considerable stability and coincidence with the principles of fatigue progression. 

\textbf{Feasibility of the input features.} RMS and MDF are traditionally used EMG features to depict fatigue, which have been demonstrated with similar characteristics with the local muscle changes of myofiber mechanical oscillations and muscular oxygen metabolism \cite{guo2022assessment}. The feature comparison demonstrates the multi-muscle properties can better depict the fatigue characteristics of walking than single muscle properties. This further indicates the muscle spatial coordination and temporal group-level neuronal activities can better reflect the fatigue-induced motor-level adaptations of complex muscle coordination, like walking.

\textbf{Weak monotonicity, trendability and suitability of fatigue.} It is common sense that as the exercise continues, the muscle fatigue accumulates and the score or the degree of fatigue should increase.  And the experimental results of fatigue \cite{sarker2020effects,gruet2013stimulation} that reveals the fatigue-induced changes of central motor regions, motor output and motor units also implicitly demonstrate the common sense and present a common trend. Moreover, the estimated fatigue score’s similarities with other views of fatigue, i.e. $Tr_{SF, F}$ and $Tr_{AD,F}$, provide further demonstrations of to what extent the estimated fatigue score follow the common trend of fatigue progression, given there is no gold standard for objective muscle fatigue, especially for multi-muscle exercise. Our comparison demonstrates the overall best performance of our method compared with the SOTA methods that can continuously assess muscle fatigue. It can be seen that our method presents similar $WO$, $Tr_{SF,F}$ compared with $WO_W$ and $WC_{M}$, and significantly better $Tr_{AD,F}$ compared with $WC_{M}$. 

\textbf{Cross-day stability.} The reason of our method’s best stability presented in our comparison can be twofold. First, the physiological rationale features extract latent states of neuro-muscular control of multi-muscle coordination \cite{chvatal2013common}, and have been demonstrated with good stability \cite{yi2021bipolar}. This can also be demonstrated by  $WC_{W}$'s and $WC_{I}$'s better suitability compared with $WC_{M}$, $WO_W$, $WO_I$ and $WO_{M}$, shown in Fig. \ref{fig:day}. Second, the proposed estimator is rationally modeled and trained by the fatigue progression-inspired loss function, thus can extract more fatigue-related information. This can be demonstrated by the better performance of our method compared with all the other SOTA estimators. even though input features can the same.

\textbf{Cross-subject and cross-view similarities}. Previous studies usually present discrete measurements of fatigue-related indices to indicate the fatigue-induced adaptations. Our proposed method provides a general measurement for multi-muscle exercise with good cross-subject similarities, compared with SOTA methods.  Moreover, fatigue studies use muscle metabolism \cite{taelman2011estimation}, muscle activation patterns \cite{fidalgo2021electromyographic}, motor region activities \cite{gruet2013stimulation} etc. to investigate the changes of specific regions under peripheral or central fatigue. And the fatigue revealed by different indexes usually presents generally similar trends over the exercise duration but different characteristics locally. In our study, the fatigue score we estimate is objective and able to reflect the fatigue progression of multiple muscles. The fatigue feelings measured by the Likert scale are subjective and have been reported to be influenced by other subjective feelings, like reward, expectation etc \cite{muller2021neural}. That accounts for the relatively low similarity between the subjective feelings and the estimated fatigue scores ($Tr_{SF,F}$). The cross-subject and cross-view similarities among the estimated muscle fatigue score, subjective fatigue feelings and the fatigue-induced shift of EMG data distribution indicate there might be a latent bodily state that regulates the fatigue-related aspects, no matter biomechanics, inner feelings or neural systems, as suggested by theoretical frameworks \cite{hureau2018sensory, weavil2019neuromuscular}. It should be noted that the fatigue-related bodily state regulation is out of the scope. We herein just propose a cue for future research.

\section{Conclusion}
This study takes the first step toward conceptualizing and continuously assessing objective muscle fatigue for daily exercise. The proposed method utilizes the features of spatial and temporal adaptations of multi-muscle coordination and group-level neuronal activities. The fatigue can then be modelled by a Bayesian Gaussian process, and trained by a time-evolving principle-inspired loss function. The experimental results demonstrate the effectiveness of our method. The promising outcomes of our study may aid fatigue monitoring, training endorsement determination and human-machine interface for more exercise modalities. 

This study is just a proof-of-concept work that has several limitations. First, the study does not consider the condition of taking rest. When there is a rest session between two exercise sessions, the monotonically increasing trend of fatigue would be violated. And the whole fatigue progression would be piece-wise monotonic. Second, more exercise modalities should be included. For example, if there is a running session after walking sessions, the increasing rate of fatigue would change and the bound of fatigue score would also break. And further running-related physiological priors can be included. Third, the model we develop utilizes the Markov, Gaussian and Bayesian assumptions. The actual fatigue progression might violate such assumptions. Future work should also develop models beyond the assumptions.

\bibliographystyle{IEEEtran}
\bibliography{IEEEabrv,IEEEexample}
\end{document}